\documentclass[11pt]{article}
\usepackage{amssymb}

\newcommand{\bra}{\begin{array}}
\newcommand{\era}{\end{array}}
\newcommand{\beq}{\begin{equation}}
\newcommand{\eeq}{\end{equation}}
\newcommand{\bqn}{\begin{eqnarray}}
\newcommand{\eqn}{\end{eqnarray}}

\def\N{\mathbb N}

\def\BC{\bb C}
\def\_\BC{\bbi C}

\def\Tr {{\rm Tr}}

\def\bz {\bar{z}}

\def\( {\left(}
\def\) {\right)}
\def\[ {\left[}
\def\] {\right]}

\def\Tr {{\rm Tr}}
\def\dag {{\dagger}}

\def\no2 {{\textstyle{n\over 2}}}

\newcommand{\te}{\theta}

\newcommand{\pa}{\partial}
\newcommand{\al}{\alpha}

\newcommand{\lb}{\label}

\newcommand{\NP}[1]{ {\it Nucl.~Phys.} {\bf #1}}

\newcommand{\PRL}[1]{ {\it Phys.~Rev.~Lett.} {\bf #1}}

\newcommand{\JMP}[1]{ {\it J. Math.~Phys.} {\bf #1}}

\begin{document}
\thispagestyle{empty}
\begin{flushright}
ucd-tpg/06-01\\
hep-th/0605290
\end{flushright}
\vspace{0.5cm}
\begin{center}
{\Large \bf Quantum Hall Droplets on Disc
and  Effective \\ Wess-Zumino-Witten Action for Edge States }\\

\vspace{0.5cm}
{\bf Mohammed Daoud$^{a,}$\footnote{ Permanent address :
Physics Department, Faculty of Sciences, University Ibn Zohr, Agadir,
Morocco.\\ e-mail : m$_{-}$daoud@hotmail.com}
and Ahmed Jellal$^{b,}$\footnote{ e-mail : ajellal@ictp.it --
  jellal@ucd.ac.ma}}\\

\vspace{0.5cm}
{\em $^a$ Max Planck Institute for The Physics of
Complex Systems,\\ N\"othnitzer Str. 38, D-01187 Dresden,
Germany}

{\em $^b$ Theoretical Physics Group, Laboratory of Condensed Matter Physics,
Faculty of Sciences,\\  Choua\"ib Doukkali University,
P.O. Box 4056,   24000 El Jadida, Morocco}
\vspace{3cm}

 {\bf Abstract}
\end{center}
\baselineskip=18pt
\medskip

We algebraically analysis the quantum Hall effect of a system of
particles living on the disc ${\bf  B}^1$ in the presence of an
uniform magnetic field $B$. For this, we identify the non-compact
disc with  the coset space  $SU(1,1)/U(1)$. This allows us to use
the geometric quantization in order to get the wavefunctions as
the   Wigner ${\cal D}$-functions satisfying a suitable
constraint. We show that the corresponding Hamiltonian coincides
with the Maass Laplacian. Restricting to the lowest Landau level,
we introduce the noncommutative geometry through the star product.
Also we discuss the state density behavior as well as the
excitation potential of the quantum Hall droplet. We show
that the edge excitations are described by an effective
Wess-Zumino-Witten action for a strong magnetic field and discuss
their nature. We finally show that
 LLL wavefunctions are intelligent states.

\newpage
\section{Introduction}

Quantum Hall effect (QHE)~\cite{prange}
has been realized on different two-dimensional manifolds.
For instance, in $1983$, Haldane~\cite{haldane} proposed an approach to overcome
the symmetry  problem that brought by the Laughlin theory~\cite{laughlin} for the
fractional QHE at the filling factor
$\nu={1\over m}$ with  $m$ is odd integer.
By considering particles living on two-sphere  ${\bf CP}^1$ in a magnetic monopole,
Haldane formulated a theory that   possess all symmetries and
generalizes the Laughlin proposal.
Very recently, Karabali and
Nair~\cite{karabali1} elaborated an elegant algebraic analysis that
supports the Haldane statement and gives a more general results.

Karabali and
Nair~\cite{karabali1} analyzed the Landau problem on   the complex projective
space ${\bf{CP}}^k$ from a theory group point of view. This analysis is based on
the fact that   ${\bf{CP}}^k$ can be  seen as the coset space
$SU(k+1)/U(k)$. More precisely, an Hamiltonian has been written in terms
of the $SU(k+1) $ generators and its spectrum has been given. Also a
link to the effective   Wess-Zumino-Witten (WZW) action for the edge states
is well established. This work has been done on the compact manifolds
and in particular two-sphere  ${\bf CP}^1$.
It is natural to ask, can we do the same analysis on non-compact
manifolds like the ball  ${\bf{B}}^k$? The present paper will partially
answer the last question and the general case will be examined
separately~\cite{daoud1}. Our motivation is based
on~\cite{karabali1} and the analytic method used by one~\cite{jellal} of the authors
to deal with QHE on  ${\bf{B}}^k$.

We algebraically investigate a system of particles living on the
disc  ${\bf{B}}^1$ in the presence of an uniform magnetic field
$B$. After realizing    ${\bf{B}}^1$ as the coset space
$SU(1,1)/U(1)$, we construct the wavefunctions as the Wigner
$\cal{D}$-functions verifying a suitable constraint. The
corresponding Hamiltonian $H$ can be written in terms of the
$SU(1,1)$ generators. This will be used to define  $H$ as a second
order differential operator, in the complex coordinates, which
coincides with the Maass Laplacian~\cite{elstrodt} on
${\bf{B}}^1$, and get the energy levels. We introduce an
excitation potential to remove the degeneracy of the ground state.
For this, we consider a potential expressed in terms of the
$SU(1,1)$ left actions. For a strong magnetic field, we show that
the excitations of the lowest landau level (LLL) are governed by
an effective Wess-Zumino-Witten (WZW) action. It turn out that
this action coincides with  one-chiral bosonic action for QHE at
the filling factor $\nu=1$~\cite{stone,iso}. We show that the
field describing the edge excitations is a superposition of
oscillating on the boundary ${\bf{S^1}}$ of the quantum Hall
droplet. Finally, we discuss the squeezing property of $SU(1,1)$
raising and lowering  operators in the lowest Landau levels and we
show that the squeezing disappear in presence of high magnetic
field.

The present paper is organized as follows.
In section 2, we present a group theory approach to analysis the Landau problem on the
disc. We build the wavefunctions and give the corresponding
Hamiltonian as well as its energy levels. In section 3, we restrict
our attention to
 LLL to write down the corresponding star product and defining the
relevant density matrix. Also we consider the excitation potential
and get the associate symbol to examine  the excited states. We
determine the effective WZW action for the edge states for a
strong magnetic field and discuss the nature of the edge states in
section 4. Section 5 is devoted to analyzing the squeezing
property of $SU(1,1)$ Weyl generators in LLL.
For this, we show that the Robertson-Schr\"odinger
uncertainty relation~\cite{rs} is minimized and according to the
literature~\cite{dod}  LLL wavefunctions are intelligent.  For a
large magnetic background, one recovers the Heisenberg uncertainty
relation and the LLL states reduces to harmonic oscillator
coherent states. We conclude and give some perspectives in the
last section.


\section{Landau problem on the disc}

We analysis the basic features of a particle living on the disc ${\bf B}^1$
in the presence of an uniform magnetic field $B$. To do this, we realize
the disc as the coset space  $SU(1,1)/U(1)$ and write down the appropriate
Hamiltonian in terms of the Casimir operator corresponding
to  $SU(1,1)$. We show that the
wavefunctions in LLL, which are obtained
to be the  Wigner ${\cal D}$-functions,  can be seen as the
coherent states of  $SU(1,1)$. These materials and related matters
will be clarified in the present section.

\subsection{Wavefunctions }

The disc is a two-dimensional non-compact surface
${\bf B}^1 =\{ z \in {\bf C}, \  \bar z\cdot z < 1\}$. The manifold  ${\bf B}^1$
can be viewed as the coset space $SU(1,1)/U(1)$
generated by the $g$ elements,
which are  $2\times
2$ matrices of a fundamental representation of the group
$SU(1,1)$. They satisfy the relation
\begin{equation}
\det g = 1,\qquad \eta g^{\dag}
\eta = g^{-1}
\end{equation}
 where $\eta= {\rm diag}(1,-1)$. An adequate parametrization can be
 written as
\begin{equation}
g= \pmatrix{\bar u_2&u_1\cr \bar u_1 & u_2\cr}
\end{equation}
where $u_1$ and $u_2$ are
the global coordinates of ${\bf B}^1$, such as
\begin{equation}
u_1 = \frac{z}{\sqrt{1-\bar z \cdot z}} {\hskip 1cm} u_2 =
\frac{1}{\sqrt{1-\bar z\cdot z}}.
\end{equation}

To generate the gauge potential, we introduce
 the Maurer-Cartan one-form $g^{-1}dg $. A straightforward
 calculation gives
\begin{equation}\lb{cmof}
g^{-1}dg = - i\ t_{+} \ e_{+} \ dz - i \ t_{-} \ e_{-}\ d\bar z - 2i\
\theta \ t_3
\end{equation}
where the one-orthonormal forms $e_{+}$ and $e_{-}$ are
\begin{equation}
e_{+} = - \frac{i}{1-\bar z\cdot z}, {\hskip 1cm} e_{-}= \frac{i}{1-\bar
z \cdot z}
\end{equation}
 and
the $U(1)$ symplectic one-form, i.e. the $U(1)$ connection, is
\begin{equation}\label{theta}
\theta = i \ {\rm Tr} \left(t_3g^{-1}dg \right) = \frac{i}{2}\
\frac{\bar z\cdot dz - z\cdot d\bar
z }{1 - \bar z\cdot z}.
\end{equation}
In (\ref{cmof}),
 $t_{+}, t_{-}$ and $t_3$
are the $SU(1,1)$ generators in the fundamental representation.
They  can be written in terms of
the   matrices $(E_{ij})_{kl} = \delta_{ik}\delta_{jl}$ of the
algebra  $gl(2)$ as
\begin{equation}
t_{+} = -E_{12},\qquad
t_{-} = E_{21},\qquad  t_3 = \frac{1}{2}(E_{11} - E_{22}).
\end{equation}
It will be clear later that (\ref{cmof}) will be used to define the
covariant derivatives in order to get a
diagonalized Hamiltonian describing the quantum system.

With the above realization, the disc
is equipped with the K\" ahler-Bergman metric
\begin{equation}
 ds^2 = \frac{1}{(1-\bar z\cdot z)^2} \ dz\cdot
d\bar z
\end{equation}
as well as a symplectic closed two-form
\begin{equation}
\omega = \frac{i}{(1-\bar z\cdot z)^2}\ dz \wedge d\bar z.
\end{equation}
One can check that the two-form is related to $\te$ as $\omega = d\theta$.
Further, $\te$ will be linked to the gauge potential of the magnetic field.
The  Poisson bracket on  ${\bf B}^1$ is given by
\begin{equation}\lb{PB}
\{f_1 , f_2 \} = i(1-\bar z\cdot z)^2\left( \frac{\partial f_1}{\partial
z}\frac{\partial f_2}{\partial \bar z} - \frac{\partial f_1}{\partial
\bar z }\frac{\partial f_2}{\partial z}\right)
\end{equation}
where $f_1$ and $f_2$ are  function on  $SU(1,1)$. They can be
expanded as
\begin{equation}
f(g) = \sum f_{m',m}^k {\cal D}_{m',m}^k (g)
\end{equation}
where the Wigner ${\cal D}$-functions ${\cal D}_{m',m}^k (g)$
on $SU(1,1)$ are
\begin{equation}\label{wdf}
{\cal D}_{m',m}^k (g) = \langle k , m'| g |k , m\rangle.
\end{equation}
$k$ is labeling the discrete  $SU(1,1)$ irreducible representation. We
choose $k$ to be integer because we are interested only to the discrete
part of the Landau system on the disk. We denote the positive discrete
representation of $SU(1,1)$ by
${D}_{+}^k$ for $2k \in {\bf N}$ and  $k>\frac{1}{2}$. For a given
$k$,   the representation space is spanned by the basis $\{
|k , m\rangle , \ m \in {\bf N}\}$.
The $SU(1,1)$ generators
\begin{equation}
\left[ t_3 , t_{\pm} \right] = \pm t_{\pm}, \qquad
 \left[ t_{-} , t_{+}  \right] = 2t_3
\end{equation}
act on the vectors basis as
\begin{equation}\label{taction}
t_{\pm} \ |k ,
m\rangle =\sqrt{\left(m+\frac{1}{2}\pm \frac{1}{2}\right) \left(2k+m-\frac{1}{2}\pm
\frac{1}{2}\right)}\ |k , m\pm 1\rangle, \qquad t_3 \ |k , m\rangle = (k+m)\ |k , m\rangle.
\end{equation}
The associated second order Casimir operator is given by
\begin{equation}
 C_2 = t_3^2-\frac{1}{2} \left(t_-t_+ +t_+t_- \right)
\end{equation}
and its  eigenvalue is $k(k-1)$.

To characterize the admissible (physical) states in analyzing the
Landau problem on  ${\bf B}^1$, we
 introduce the generators of the right $R_a$ and left $L_a$
translations of $g$
\begin{equation}\lb{lrgerators}
R_a g = g t_a,\qquad  L_a g = t_a g
\end{equation}
where  $ a $ runs for $= +, -, 3$. They act on the Wigner ${\cal D}$-functions as
\begin{equation}
R_a {\cal D}_{m',m}^k (g) = {\cal D}_{m',m}^k (gt_a), \qquad
L_a {\cal D}_{m',m}^k (g) = {\cal D}_{m',m}^k (t_ag).
\end{equation}

To obtain the Hilbert space corresponding to the quantum
system living on ${\bf B}^1$,
we should reduce the
degrees of freedom on the manifold $SU(1,1)$ to
the coset space $SU(1,1)/U(1)$.  It will be clear soon that this reduction can be formulated
in terms of a suitable constraint on the Wigner ${\cal D}$-functions. Note that, the present system is
submitted to the magnetic strength
\begin{equation}
F = dA
\end{equation}
where the $U(1)$ gauge field potential is given by
\begin{equation}
A = n \frac{i}{2}\ \frac{\bar z \cdot dz - z\cdot d\bar z }{1 - \bar z\cdot z}.
\end{equation}
where $n$ is a real number. It is obvious that $A$ is proportional
to the one-form (\ref{theta}), i.e. $A=n\te$. Since we have a
closed two-form $\omega = d\theta$, the components of the magnetic
field expressed in terms of the frame fields defined by the metric
are constants. Hereafter, we set
\beq
n = \frac{B}{2}.
\eeq

The suitable constraint on the Wigner ${\cal D}$-functions can be
established by considering the $U(1)$ gauge transformation
\begin{equation}\lb{gtran}
g \rightarrow gh = g\exp(it_3\varphi)
\end{equation}
where $\varphi$ is the $U(1)$ parameter. (\ref{gtran})
leads  to the transformation in the gauge field
\begin{equation}
A \rightarrow A + nd\varphi.
\end{equation}
It follows that the functions~(\ref{wdf}) transform as
\begin{equation}
{\cal D}_{m',m}^k (gh) = \exp\left(\int\dot{A}dt\right){\cal D}_{m',m}^k (g)
= \exp\left(\frac{n}{2}\varphi\right){\cal D}_{m',m}^k (g).
\end{equation}
Therefore, the canonical momentum corresponding to the $U(1)$ direction
is $n/2$. Thus, an admissible quantum states $\psi\equiv{\cal D}_{m',m}^k(g)$ must satisfy the
constraint
\begin{equation}\label{constraint}
R_3 \psi = \frac{n}{2} \psi.
\end{equation}
Equivalently, we have
\begin{equation}\lb{rcaction}
[ R_- , R_+ ] = n.
\end{equation}
This relation is very interesting in many respects. Indeed, the
operators $R_+$ and $ R_-$ can be seen
as the creation and annihilation operators in analogy with the standard
harmonic oscillator involved in the Landau problem on the plane. A similar
result was obtained by Karabali and Nair~\cite{karabali1} in dealing
with the same problem on  ${\bf CP}^1$. Furthermore,~(\ref{rcaction}) is suggestive
to make  contact with
the noncommutative geometry
through the star product.
Finally, the physical states
constrained by~(\ref{constraint}) are  the Wigner ${\cal D}$-functions ${\cal D}_{m',m}^k
(g)$ with the condition
\beq
m = \frac{n}{2}-k.
\eeq

The
Lowest Landau condition is
\begin{equation}\label{cond}
R_- {\cal D}_{m',m}^k (g) = 0
\end{equation}
which corresponds to $m=0$, i.e. $k=\frac{n}{2}$. This shows that the
index $k$, labeling
the $SU(1,1)$ irreducible representations, is related to the
magnetic field by $k = \frac{B}{4}$. In the geometric
quantization language~(\ref{cond}) is called the polarization
condition. This
is an holomorphicity  condition, which means that the LLL wavefunctions
${\cal D}_{m',0}^k (g)$
\begin{equation}
\psi_{LLL} \equiv {\cal D}_{m',0}^k (g)= \langle k,m'|g|k,0\rangle
\end{equation}
are  holomorphic in the $z$ coordinate. More precisely,
in the fundamental representation,  we can define $g$
in terms of the generators $t_{\pm}$ by
\begin{equation}\lb{gexp}
g = \exp(\eta t_+ - t_-\bar \eta)
\end{equation}
where $\eta$ is related to the local coordinates via
\begin{equation}
z = \frac{\eta}{|\eta|}\tanh|\eta|.
\end{equation}
Using (\ref{taction}), we end up with the required wavefunctions
\begin{equation}\lb{lllf}
\psi_{LLL} (\bz, z) = (1-\bar
z\cdot z)^{\frac{n}{2}}\sqrt{\frac{(m'+n-1)!}{m'!(n-1)!}}z^{m'}, \qquad m' \in {\bf N}.
\end{equation}
Note that, LLL is
infinitely degenerated and $\psi_{LLL}$ are nothing but the $SU(1,1)$ coherent
states. They constitute an over-complete set with respect to the
measure
\begin{equation}\lb{mesure}
d\mu( \bar z, z) = \frac{n-1}{\pi}\ {d^2z\over (1-\bar z\cdot z)^{2}}.
\end{equation}
The orthogonality relation writes as
\begin{equation}
\int d\mu(\bz, z)\ {\cal D}_{m",0}^{\star
k} (g)\ {\cal D}_{m',0}^k (g) = \delta_{m',m"}.
\end{equation}
Recall that the LLL wavefunctions
of a particle living on two-sphere coincide with the $SU(2)$ coherent states 
\cite{karabali1}.
Note also that for the Landau problem on the plane, the LLL vectors are given by
the harmonic oscillator coherent states \cite{iso}. 

At this level, it is natural to look for the energy levels
corresponding to the
wavefunctions  ${\cal D}_{m',{n\over 2}-k}^k (g)$.
These can
be obtained  by defining the relevant  Hamiltonian that describes
the quantum system living on the coset space $SU(1,1)/U(1)$.

\subsection{Hamiltonian and energy levels }

To derive the appropriate Hamiltonian, we start by noting that
from above and more precisely relation~(\ref{rcaction}), $R_+$ and
$R_-$ can be seen, respectively,
as raising and lowering operators. This is in analogy with the
creation and annihilation operators corresponding to the standard
harmonic oscillator. Therefore, the Hamiltonian, describing a system of
charged particle living on the disc in the presence of a background
field, can be written as
\begin{equation}
H = \frac{1}{2} \left(R_-R_+ + R_+R_-\right).
\end{equation}
To write this Hamiltonian in terms of the complex coordinates $z$
and $\bz$, we introduce the
$U(1)$ covariant derivatives. These can be obtained
from~(\ref{cmof}), such as
\begin{equation}
D_z = \frac{\partial}{\partial z} - i A_z, \qquad  D_{\bar z} =
\frac{\partial}{\partial \bar z} - i A_{\bar z}
\end{equation}
where the components of the gauge potential have the forms
\begin{equation}
A_z = i\frac{n}{2} \frac{\bar z}{1 - \bar zz}, \qquad
A_{\bar z} =-i\frac{n}{2} \frac{z}{1 - \bar zz}.
\end{equation}
Using (\ref{cmof}) and (\ref{lrgerators}), we can map
the raising  and lowering operators $R_{\pm}$ in terms of  $D_{z}$ and
$D_{\bz}$ as
\begin{equation}
R_+ =- \left(1 - \bar z\cdot z \right) D_z, \qquad R_- = \left(1 -
\bar z\cdot z
\right)
D_{\bar z}.
\end{equation}
With these relations,  we finally end up with  the second order differential
realization of the required Hamiltonian. This is
\begin{equation}
H = -(1 - \bar z\cdot z)\left\{(1-\bar z\cdot z)\frac{\partial}{\partial z
}\frac{\partial}{\partial \bar z} +
\frac{n}{2}\bigg(z\frac{\partial}{\partial z }-\bar
z\frac{\partial}{\partial \bar z}\bigg)\right\} + \frac{n^2}{4}\bar z\cdot z.
\end{equation}
This exactly coincides with the Maass Laplacian~\cite{elstrodt}. It
 has been investigated at many occasions and generalized to the
higher dimensional spaces~\cite{groupm}.

To establish a relation between  the Casimir operator
and the Hamiltonian, we may write the eigenvalue
equation as
\begin{equation}\lb{hampo}
  H \psi = \frac{1}{2}\left(t_-t_+ + t_+t_- \right)\psi = E \psi.
\end{equation}
Since the wavefunctions $\psi$ are the Wigner ${\cal D}$-functions ${\cal
  D}_{m',{n\over 2}-k}^k (g)$, the Landau energies are given in terms of $C_2$
\begin{equation}\lb{spec}
E =  \frac{n^2}{4} - C_2
\end{equation}
which gives
\begin{equation}
E_m = {n\over 2}\left(2m+1\right)- m\left(m+1 \right).
\end{equation}
This is similar to that obtained in~\cite{jellal} and references
therein. It is clear from (\ref{hampo}) that the eigenvalues of $H$ must be
positive. This implies the constraint
$0\leq m<{n-1\over 2} $ and therefore we have a finite number of 
Landau levels, each level is infinitely degenerated. $E_m$ can be 
campared to one particle spectrum on the sphere~\cite{karabali1} 
\begin{equation}
E_m^{\sf 
 sphere} = {n\over 2}\left(2m+1\right)+ m\left(m+1 \right),\qquad 
m\in {\N}
\end{equation}
where the degeneracy of the Landau levels is
finite. Note that, for large $n$
 we get the
Landau spectrum on the Euclidean surface
\begin{equation}
E_m^{\sf 
 plane} = {n\over 2}\left(2m+1\right), \qquad 
m\in {\N}
\end{equation}
showing that the landau levels are infinitely degenerated.
 The energy levels are
indexing by the integer $m$ and for $m=0$ we obtain the LLL energy
\begin{equation}
E_0 = \frac{n}{2} = \frac{B}{4}.
\end{equation}
This coincides with the ground state of the same problem on the plane.
It will be investigated carefully to make contact with QHE at LLL
on the coset space ${SU(1,1)/U(1)}$.

\section{Lowest Landau level analysis}

For our end, we establish some relevant ingredients.
 These are the star product, density matrix and
excitation potential. We will see how these will play a crucial role
in determining the effective WZW action describing the edge states for
 a strong magnetic
field and discussing the nature of the edge excitations.

\subsection{Star product}

To derive the effective action
describing the edge excitations, we will replace the commutators of two
operators by a non-commutative Moyal bracket. It
coincides  with the Poisson  bracket for a strong magnetic field. Recall
that for large $n$ $(B\sim n)$, the particles are constrained to be confined in
LLL described by~(\ref{lllf}).

To define the star product in terms of our language, we start by
noting that for any operator ${A}$ acting on $\psi_{LLL}$, we can
associate the function
\begin{equation}
{\cal A}(\bar z, z) = \langle z | \hat{A} | z \rangle = \langle 0
|g^{\dag} {A} g| 0 \rangle
\end{equation}
where $g$ is given by~(\ref{gexp}) and $| 0 \rangle \equiv |k,0\rangle$ is the
lowest highest weight state  of the discrete $SU(1,1)$ representation.
The vector states $| z \rangle =g| 0 \rangle $ are the $SU(1,1)$ coherent states
\begin{equation}\lb{cs}
| z \rangle  = (1-\bar
z\cdot z)^{\frac{n}{2}}\sum_{m'=0}^{\infty}\sqrt{\frac{\left(m'+n-1\right)!}{m'!(n-1)!}}z^{m'}|k,m'\rangle.
\end{equation}
Note that $\psi_{LLL}$ is nothing but the projection of $| z \rangle $
on the state $|k,m'\rangle $
\begin{equation}
\psi_{LLL} = \langle k, m'|z\rangle.
\end{equation}
These can be used to define
an associative star product of two functions ${\cal A}(\bar z, z)$ and ${\cal
B}(\bar z, z)$ by
\begin{equation}\lb{sp2f}
{\cal A}(\bar z, z)\star {\cal B}(\bar z, z) = \langle z | {A} {B}  | z
\rangle.
\end{equation}
With the help of  the unitary condition of $g$, i.e. $ g^{\dag}g = 1$, and the completeness
relation
\begin{equation}
\sum_{m=0}^{\infty} \ |k,m\rangle \langle k,m|= 1
\end{equation}
we are able to write
\begin{equation}
{\cal A}(\bar z, z)\star {\cal B}(\bar z, z) = \sum_{m=0}^{\infty}
\langle 0 |g^{\dag} {A} g |k,m\rangle \langle k,m|g^{\dag}  {B} g | 0
\rangle.
\end{equation}
Using~(\ref{taction}) and  at large $n $, we show that the star product is
\begin{equation}\lb{sp2f2}
{\cal A}(\bar z, z)\star {\cal B}(\bar z, z) = {\cal A}(\bar z, z)
{\cal B}(\bar z, z) + \frac{1}{n} \langle 0 |g^{\dag} {A} g t_{+}
|0\rangle \langle 0|t_{-}g^{\dag} {B} g| 0 \rangle +
O \left(\frac{1}{n^2} \right).
\end{equation}
It is clear that the first term in h.r.s. is the
ordinary product of two functions ${\cal A}$ and ${\cal B}$. While, the
non-commutativity is encoded in the second term.
This is interesting result, because it will play an important role
when we construct the effective action for the edge states.

 To obtain the final
form of the star product~(\ref{sp2f2}), it is necessary to evaluate
the matrix element of type
\begin{equation}
\langle 0 |g^{\dag} A g t_{+i}
|0\rangle.
\end{equation}
To do this task, we can use the coherent states~(\ref{cs}) to write
the holmorphicity condition as
\begin{equation}
R_{-}\langle k,m|g |0\rangle = 0.
\end{equation}
Thus, we have
\begin{equation}
 \langle 0 |g^{\dag} {A}  g t_{+}
|0\rangle = R_{+} \langle 0 |g^{\dag} {A}   g  |0\rangle.
\end{equation}
Similarly, we obtain
\begin{equation}
 \langle
0|t_{-}g^{\dag} B g| 0 \rangle = - R_{-}\langle 0|g^{\dag} B g| 0
\rangle
\end{equation}
where we have used the condition $R_{+i}^{\star} = - R_{-i}$. 
From the above equations and
since we are concerned with a  $U(1)$ abelian gauge
field, we show that the star product~(\ref{sp2f2}) becomes
\begin{equation}\lb{sp2f3}
{\cal A}\left(\bar z, z\right)\star {\cal B}\left(\bar z, z\right) = {\cal A}\left(\bar z, z\right)
{\cal B}\left(\bar z, z\right) - \frac{1}{n}\left(1 - \bar zz\right)^2
\partial_z {\cal A}\left(\bar z, z\right)\partial_{\bar z}{\cal B}\left(\bar z,
z\right) + O \left(\frac{1}{n^2} \right).
\end{equation}
Therefore, the symbol or function associated to the commutator of two
operators ${A}$ and ${B}$ can be written as
\begin{equation}\lb{sp2f4}
\langle z |\left[ A , B\right] | z \rangle = - \frac{1}{n}\left(1 - \bar zz\right)^2
\left\{\partial_{z}{\cal A}\left(\bar z, z\right)\partial_{\bar z}{\cal B}\left(\bar z,
z\right) - \partial_{z}{\cal B}\left(\bar z, z\right)\partial_{\bar z}{\cal A}\left(\bar
z, z\right)\right\}.
\end{equation}
This implies
\begin{equation}\lb{sp2f5}
\langle z |\left[ A , B\right] | z \rangle =  \frac{i}{n} \left\{{\cal A}\left(\bar z,
z\right), {\cal B}\left(\bar z, z\right)\right\} \equiv \left\{{\cal A}\left(\bar z, z\right), {\cal
B}\left(\bar z, z\right)\right\}_{\star}
\end{equation}
where $\{ , \}$ stands for the Poisson bracket~(\ref{PB}) on the disc and
$\{ , \}_{\star}$ for the
Moyal bracket defined by
\beq
\left\{{\cal A}\left(\bar z, z\right), {\cal B}\left(\bar z, z\right)\right\}_{\star} = {\cal
A}\left(\bar z, z\right)\star {\cal B}\left(\bar z, z\right) - {\cal B}\left(\bar z, z\right)\star
{\cal A}\left(\bar z, z\right).
\eeq

The advantage of the obtained star product
will be seen in the construction of the effective WZW action
describing the edge excitations of the the quantum Hall droplet.

\subsection{Density matrix}

Another important ingredient that should be investigated is the density
matrix.
Note that, in the disc, LLL are infinitely degenerated and
one may
fill the LLL states with $M$ particles, $M$ very large. The corresponding density operator
is
\begin{equation}\lb{dens}
{\rho}_0 = \sum_{m=0}^M| k,m \rangle \langle k,m |.
\end{equation}
The associated symbol can be written as
\begin{equation}
\rho_0(\bar z, z) = (1-\bar z\cdot z)^n \sum_{m = 0}^{M}
\frac{(n-1+m)!}{(n-1)!m!}(\bar z\cdot z)^{2m}.
\end{equation}

To analyze the behavior of  $\rho_0(\bar z, z) $ for a strong
magnetic field, we note that the normalization factor can be
expanded as
\begin{equation}
(1-\bar z\cdot z)^{-n} =  \sum_{m}^{\infty}
\frac{(n-1+m)!}{(n-1)!m!}(\bar z\cdot z)^{2m}
\end{equation}
which gives for large $n$
\begin{equation}
(1-\bar z\cdot z)^{n} =  \exp(-n\bar z\cdot z).
\end{equation}
On the other hand, one can see
that the second term occurring in
the expression of $\rho_0(\bar z, z)$ behaves for large $n$ as the following series
\begin{equation}
\sum_{m=0}^M \frac{(n\bar z\cdot z)^m}{m!}.
\end{equation}
Combining all, we obtain an approximated density for large $n$
\begin{equation}\lb{lnd}
\rho_0(\bar z, z) \simeq \exp\left(-n\bar z\cdot z\right)\sum_{m=0}^M \frac{(n\bar
z\cdot z)^m}{m!}\simeq \Theta \left(M - n\bar z\cdot z\right).
\end{equation}
This expression is valid for a large number $M$ of
particles~\cite{sakita}.
The mean value of the density operator, in LLL, is a step
function for $n\rightarrow \infty$ and $M\rightarrow \infty$
($\frac{M}{n}$ fixed). It corresponds to an abelian droplet
configuration with boundary defined by
\beq\lb{boun}
n\bar z\cdot z = M
\eeq
 and its
radius is proportional to $\sqrt{M}$. Furthermore, the derivative of the
density $\rho_0(\bar z, z)$ tends to a $\delta$-function. This property will be useful
in the description of the edge excitations.

\subsection{Excitation potential}

Once we determined the spectrum of LLL where the quantum Hall droplet 
is specified by the density matrix $\rho_0$, one may ask about the
excited states. 
The answer can be given by describing the excitations in terms of an
unitary time evolution operator $U$. It
contains  information concerning the dynamics of the excitations
around $\rho_0$. Therefore the excited states will be characterized by a
density operator given by
\beq\lb{rho}
{\rho} = U {\rho}_0 U^{\dag}.
\eeq
This is basically corresponding to a perturbation of the quantum system.
Its relevant Hamiltonian can be written as
\begin{equation}\lb{ch}
{\cal H} = E_0 + V
\end{equation}
where $E_0 = \frac{n}{2}$ is the LLL energy and $V$ is the
excitation potential. This perturbation  will induces  a lifting
of the LLL degeneracy. Note that, the $SU(1,1)$ left actions commute
with the covariant derivatives. They correspond to the magnetic
translations on the disc and lead to degeneracy of the Landau
levels. Thus, it is natural to assume that  $V$
as a function of the magnetic translations $L_3, L_+$ and
$L_-$. A simple choice for the potential $V$ is
\begin{equation}\lb{evff}
V = \omega \left(L_3 - \frac{n}{2} \right).
\end{equation}
The symbol associated
to this potential is given by
\begin{equation}\lb{evf}
 {\cal V}(\bar z, z ) = \langle z |V| z \rangle = n\omega \frac{\bar z\cdot z}{1-\bar z\cdot z}.
\end{equation}
It goes essentially  to the harmonic oscillator potential for a strong
magnetic field. One now can verify that the spectrum of~(\ref{ch}) is
\begin{equation}
{\cal H} \psi_{LLL}\equiv {\cal H} {\cal D}_{m',0}^k (g)=
\left(E_0 + \omega m'\right){\cal D}_{m',0}^k (g).
\end{equation}
This shows that we have a lifting of the LLL degeneracy.

\section{Wess-Zumino-Witten action for edge states}

The analysis developed in the previous section is useful
 to derive the effective WZW action for the edge
states. Recall that for a  strong magnetic field, the particles are
confined in LLL. The required action will basically describe the
 behavior of the quantum system on LLL.

\subsection{Effective action}

As mentioned above, the
dynamical information related to the degrees of freedom of the edge
states, is contained in the unitary operator $U$~(\ref{rho}). The action,
describing these excitations, in the Hartree-Fock approximation,
can be written as~\cite{sakita1}
\begin{equation}\lb{effa}
S = \int dt\ \Tr \ \left( i\rho_0 U^{\dag} \partial_t U - \rho_0 U^{\dag}{\cal H}U \right)
\end{equation}
where ${\cal H}$ is given by~(\ref{ch}). For a strong magnetic
field, i.e. large $n$, the different quantities occurring in the action can be
evaluated as classical functions. To do this, we adopt a
method similar to that used in~\cite{karabali1}. This is mainly based on the strategy
elaborated by Sakita~\cite{sakita1} in dealing with a bosonized theory for
fermions.

To determine the effective action,
we start by calculating the first term
in r.h.s. of~(\ref{effa}). This can be done by setting
\beq
U= e^{+i\Phi}, \qquad
\Phi^{\dag} = \Phi.
\eeq
 This suggests to write $dU$ as
\begin{equation}
dU = \sum_{n=1}^{\infty}\frac{(i)^n}{n!}\sum_{p=0}^{n-1}\ \Phi^p
\ d\Phi \ \Phi^{n-1-p}
\end{equation}
as well as the operator $U^{\dag}dU $
\begin{equation}
U^{\dag}dU = i \int_0^1 d\alpha \ e^{-i\alpha\Phi} \ d\Phi \
e^{+i\alpha\Phi}.
\end{equation}
This leads to the relation
\begin{equation}
e^{-i\Phi}\ \partial_t \ e^{+i\Phi} = i \int_0^1 d\alpha \ e^{-i\alpha
 \Phi} \ \partial_t\Phi \ e^{+i\alpha\Phi}.
\end{equation}
We show that  the first term in r.h.s. of~(\ref{effa}) is
\begin{equation}\lb{1rhs}
 i \int dt \ \Tr\left(\rho_0 U^{\dag}\partial_tU \right) =
- \sum_{n=0}^{\infty} \frac{ (i)^n}{(n+1)!} \
\Tr
\left(\underbrace{[\Phi,\cdots[\Phi}_n,\rho_0]\cdots]\partial_t\Phi
\right).
\end{equation}
Due to the completeness  of  LLL, the trace of any
operator ${A}$ is  defined by
\beq
\Tr {A} = \int d\mu(z, \bar z )\ \langle z | {A} | z
\rangle
\eeq
where the measure $d\mu(z, \bar z )$ is given by~(\ref{mesure}).
It follows that~(\ref{1rhs}) can be written as
\begin{equation}\lb{bf1t}
 i \int dt\ \Tr\left(\rho_0
U^{\dag}\partial_tU\right) =-\int d\mu(z, \bar z )\ \sum_{n=0}^{\infty}
\frac{ (i)^n}{(n+1)!} \
\underbrace{\{{\phi},\cdots\{\phi}_n,\rho_0\}_{\star}\cdots\}_{\star}\star\partial_t\phi
\end{equation}
with $\phi =\langle z| \Phi|z\rangle$.
This form is more suggestive for our purpose. Indeed, using the relations
(\ref{sp2f3}-\ref{sp2f5}), it is easy to see that~(\ref{bf1t})
rewrites as
\begin{equation}
 i \int dt \ \Tr\left(\rho_0 U^{\dag}\partial_tU \right) \simeq \frac
 {1}{2n} \int d\mu( \bar z, z )
\ \{\phi,\rho_0\} \ \partial_t\phi
\end{equation}
where we have dropped the terms in $\frac{1}{n^2}$ 
as well as the total time derivative. 
The Poisson bracket can be calculated to get
\begin{equation}\lb{epb}
\{\phi , \rho_0\} = ({\cal L}\phi)  \frac{\partial\rho_0}{\partial
(\bar z\cdot z)}
\end{equation}
and  the first order differential operator ${\cal L}$ is given by
\begin{equation}\lb{amomenta}
 {\cal L} =  i \left(1 - \bar z\cdot z\right)^2 \left(z\cdot \frac{\partial}{\partial z}
- \bar z\cdot \frac{\partial}{\partial \bar z}\right).
\end{equation}
This is the angular momenta mapped in terms of the local coordinates
of the disc.
Recall that, for large $n$, the density~(\ref{lnd}) is a step
function. Its derivative is a $\delta$-function with a support on
the boundary $\partial {\cal D}={\bf S}^1$ of the quantum Hall droplet ${\cal D}$
defined by~(\ref{boun}). By setting $z = re^{i\al}$,
we show that~(\ref{amomenta}) reduces to $ {\cal L} =
\partial_{\al}$ for large $n$. Therefore, the equation
(\ref{amomenta}) takes the form
\begin{equation}\lb{fterm}
 i \int dt\ \Tr\left(\rho_0 U^{\dag}\partial_tU \right) \approx
 -\frac{1}{2} \int_{{\bf S^1}\times{\bf
R^+}} dt \ \left(\partial_{\al}\phi\right) \left(\partial_t\phi \right).
\end{equation}

To achieve the derivation of edge states action,
it remains to evaluate the second term in r.h.s. of~(\ref{effa}).
By a straightforward calculation, we find
\begin{equation}\lb{sterm}
\Tr\left(\rho_0 U^{\dag} V U\right) = \Tr\left(\rho_0  V \right) + i
\Tr\left(\left[\rho_0, V\right] \Phi\right)
+ \frac{1}{2}\  \Tr\left(\left[\rho_0, \Phi \right] \left[V, \Phi
  \right] \right).
\end{equation}
The first term in r.h.s of~(\ref{sterm}) is $\Phi$-independent. We can drop
it because does not contain any information about the dynamics of the
edge excitations. While, the second term can be written in term of the
Moyal bracket as
\begin{equation}
 i \Tr\left(\left[\rho_0, V\right] \Phi\right) \approx i
\int d\mu(\bar z, z)\ \left\{\rho_0, {\cal V}\right\}_{\star}\ \phi.
\end{equation}
Using~(\ref{evf}), one can see that
\begin{equation}
 i \Tr\left(\left[\rho_0, V\right] \Phi\right)  \longrightarrow 0
\end{equation}
The last term in r.h.s of~(\ref{sterm}) can be evaluated in a similar
way to get~(\ref{epb}). Therefore, adding different
terms, we obtain
\begin{equation}\lb{sterm2}
  \int dt\ \Tr\left(\rho_0
U^{\dag}{\cal H}U\right) = - \frac{1}{2n^2}\int d\mu(\bar z, z)\ \left({\cal L
}\phi\right)\ \frac{\partial\rho_0}{\partial (\bar z\cdot z)}\ \left({\cal L}\phi\right)
\ \frac{\partial {\cal V}}{\partial (\bar z\cdot z)}.
\end{equation}
Note that, we have eliminated a term containing the ground state
energy $E_0$, because does not contribute to the edge dynamics.
For large $n$, from~(\ref{evf}), we notice that
\beq
\frac{\partial {\cal V}}{\partial (\bar z\cdot z)} \longrightarrow
n\omega.
\eeq
Using the spatial shape of density $\rho_0$, we
finally obtain
\begin{equation}\lb{sterm3}
 \int dt\ \Tr(\rho_0
U^{\dag}{\cal H}U) = \frac{\omega}{2}\int_{{\bf S^1}\times {\bf R^+}}
dt \ \left(\partial_{\al}\phi\right)^2.
\end{equation}
Combining~(\ref{fterm}) and~(\ref{sterm3}), we find the
appropriate effective action
\begin{equation}\lb{feaction}
S \approx -\frac{1}{2}\int_{{\bf S^1}\times {\bf R^+}} dt \
\left\{\left(\partial_{\al}\phi\right) \ \left(
\partial_t\phi \right)+\omega \ \left(\partial_{\al}\phi\right)^2\right\}.
\end{equation}
This action is actually
 describing the edge excitations of the quantum Hall
droplet. It involves only the time derivative $\pa_t\phi$ and the
tangential derivative $\partial_{\al}\phi$. The action~(\ref{feaction}) coincides
with the well-known one-dimensional chiral bosonic action
describing the edge excitations for QHE at the filling factor $\nu = 1$~\cite{stone,iso}.
This is one of the most important results derived in the present
paper.

\subsection{Nature of edge excitations}

Starting from the action~(\ref{feaction}),
we discuss the nature of the edge excitations.
This can be done by solving the
equation of motion for the field $\phi$
\begin{equation}\lb{eom}
\partial_{\al}\left( \partial_t\phi + \omega
\partial_{\al}\phi\right)= 0.
\end{equation}
The general solutions are
\begin{equation}
\phi(\alpha , t) = \phi_1(\alpha - \omega t) + \phi_2(t),
\end{equation}
which look like the right-moving waves, but in addition
there is a hidden gauge symmetry encoded in the term $\phi_2(t)$.
It corresponds to the invariance of the
action~(\ref{feaction}) under the change
\beq
\phi \rightarrow \phi + \lambda(t).
\eeq
This takes its origin from the invariance under the
$U(1)$ transformation, such as
\beq
U \rightarrow \exp\left[i\lambda(t)\right]U.
\eeq
As far as the coset space $SU(1,1)/U(1)$ is concerned, $\phi_2(t)$ does
not represent any physical degree of freedom. It can be removed by
imposing the gauge constraint
\begin{equation}
\left( \partial_t + \omega
\partial_{\al}\right)\phi= 0.
\end{equation}
Since the edge excitations action~(\ref{feaction}) is defined on the
boundary ${\bf S}^1$, we
also impose the boundary condition
\begin{equation}
\phi(2\pi,t) - \phi(0,t)= - 2\pi q
\end{equation}
and $q$ is a time independent constant. The general form of the
field $\phi(\alpha,t)$ is then given by
\begin{equation}\lb{phiphi}
\phi(\alpha - \omega t)= p - q(\alpha - \omega t) + i \sum_{n \ne
0}^{\infty}\frac{\alpha_n}{n}e^{in(\alpha - \omega t)}
\end{equation}
where $p$ is the
canonical momentum and we have set  $\alpha_{-n} = \alpha_n^{\star}$. The canonical
momentum corresponding to the field $\phi$ is
\begin{equation}
\pi(\alpha,t) =  q +  \sum_{n\ne 0} \alpha_n e^{in(\alpha - \omega
t)}.
\end{equation}
The quantization of the theory can be performed by imposing the
equal time canonical commutation relation, such as
\begin{equation}
[\pi(\alpha,t), \phi(\alpha',t)] = i \delta(\alpha - \alpha').
\end{equation}
This implies that $p$, $q$ and $\alpha_n$ must satisfy the algebra
\begin{equation}
\left[\alpha_n, \alpha_m\right] =  \delta_{n+m,0}, \qquad  \left[ q , p\right] = i
\end{equation}
and other commutators vanish. This algebra is describing an
infinite set of uncoupled oscillators. It shows that the field
$\phi$ is a superposition of oscillating modes on the boundary
$\bf{S^1}$. In other words, the edge excitations constitute the
low-lying excitations about the incompressible Hall droplet. In this
case,
the Hilbert space is the product of the oscillator Fock
spaces. The Hamiltonian of the edge excitations is given,
 with an appropriate normal ordering, by
\begin{equation}
H_e = \frac{1}{4\pi} \int_{{\bf S}^1} :(\partial_{\alpha}\phi)^2:.
\end{equation}
It can also be written as
\begin{equation}
H_e = \frac{1}{2}\alpha_0^2 + \sum_{n>0} \alpha_n\alpha_{-n}
\end{equation}
where we have $\alpha_0 = q$. At this level, it is interesting to note
that the charge operators $L_0 = H_e$ and
\begin{equation}
L_n = \frac{1}{4\pi} \int_{0}^{2\pi}
:(\partial_{\alpha}\phi)^2:e^{-in(\alpha - \omega t)} =
\frac{1}{2} \sum_{l=-\infty}^{+\infty}\alpha_{n-l}\alpha_{n},
{\hskip 1cm} n\ne 0
\end{equation}
satisfy the Virasoro algebra
\begin{equation}
[L_n , L_m] = (n-m) L_{n+m} +\frac{1}{12}(n^3 - n)\delta_{n+m , 0}
\end{equation}
of central charge $c = 1$. This show that the dynamics of the edge
excitations are governed by $(1+1)$ conformal field theory.
This link
were studied in~\cite{wen,cap} in order to describe the excitations
on the boundary of the quantum Hall droplets. Although, the result
obtained in~\cite{cap}  concerned the disc geometry, it is mainly
based on the analysis of the LLL wavefunctions of a non-relativistic
particle living on the plane, which are the harmonic
oscillator coherent states. In our case, the relation between the edge
dynamics and the non-relativistic particles on ${\bf B}^1$ in LLL
is shown up  by deriving the WZW action. This
derivation based on the  $SU(1,1)$ coherent states and the notion of star
product to evaluate the transition amplitude under the confining
potential
$V$ given by~(\ref{evff}).

The space, generated by the zero modes, plies an important role in
QHE, especially when the filling factor is
fractional. Indeed,  the edge states for the fractional Hall
effect at the filling factor $\nu =\frac{1}{2m+1}$, with $m$ integer,
  can be described by
the obtained  WZW action. It can be simply obtained by
substituting
\beq
p\rightarrow \frac{1}{\sqrt{2m+1}}p, \qquad
q\rightarrow \sqrt{2m+1}q
\eeq
 in the expression of $\phi(\alpha -
\omega t)$~(\ref{phiphi}). The corresponding Hamiltonian takes the
form
\begin{equation}\lb{wham}
H_e = \frac{1}{2}(2m+1)\alpha_0^2 + \sum_{n>0} \alpha_n\alpha_{-n}.
\end{equation}
This is exactly the Hamiltonian analyzed by Wen~\cite{wen} in
describing the edge excitations. It is interesting to note that, in a strong magnetic
field,  (\ref{wham})
provides a description of a system of anyons, whose
statistical parameter is $2m$~\cite{Hans}. It follows that the
obtained WZW action  can
also be used to describe particles with intermediate statistics.
This is essentially due to the fact that QHE as well as anyon
systems in two-dimensions are involved the $(2+1)$ Chern-Simons
interaction.

\section{Uncertainty relation in LLL}

We show that the LLL wavefunctions are intelligent, i.e.
they
minimize the
Robertson-Schr\"odinger (RS) uncertainty relation~\cite{rs}, for instance
see also~\cite{dod}.
We also show that the correlation
between the left $SU(1,1)$ Weyl generators vanishes in for a strong
magnetic field and the RS uncertainty
relation reduces to Heisenberg one. In this case, the LLL wavefunctions
behave like those corresponding to the standard harmonic oscillators
and the underlying
dynamical algebra reduces to the Weyl-Heisenberg one. The latter
provides us with a realization of fuzzy disc in the LLL. It seems
that the fact of approximating the algebra of functions on the disc by
a finite dimensional matrix model is related to the absence of
correlation between the raising and lowering operators on LLL.

To show that the the states~(\ref{cs}) saturate the RS
uncertainty relation, we start by evaluating  the mean values of the
generators $L_{\pm} = L_1 \pm i L_2$ and $L_3$. These are
\begin{equation}
\langle L_- \rangle = \overline{\langle L_+ \rangle} = n
\frac{z}{1-\bar z.z} {\hskip 1cm} \langle L_3 \rangle =
\frac{n}{2} + {\cal V}(\bar z, z)
\end{equation}
where ${\cal V}(\bar z, z)$ is the function associated to
excitation potential given by~(\ref{evf}). It follows that the
dispersion of the generators $L_j$ writes as
\begin{equation}
\sigma_j^2 = \langle L_j^2 \rangle - \langle L_j \rangle^2 =
\frac{n}{4} \frac{\vert 1 - (-)^j z^2\vert^2}{(1- \bar z.z)^2}, \qquad j =1,2.
\end{equation}
The correlation of $L_1$ and $L_2$ is given by
\begin{equation}
\sigma_{12} = \frac{1}{2}\langle L_1L_2 +  L_2L_1 \rangle  -
\langle L_1 \rangle \langle L_2 \rangle= i \frac{n}{4} \frac{z^2 - \bar
z^2}{(1- \bar z.z)^2}
\end{equation}
It is now easy  to check that
the LLL wavefunctions minimize the RS uncertainty relation
\begin{equation}
\sigma_1^2 \sigma_2^2 =  \frac{1}{4} \langle L_3 \rangle^2 +
\sigma_{12}^2
\end{equation}
and therefore they are intelligent. For a large magnetic field,
this relation reduces to Heisenberg one. 
We verify
that the correlation
\begin{equation}
\sigma_{12} \sim O\left(\frac{1}{n}\right)
\end{equation}
vanishes for $n$ large and the RS uncertainty relation gives the
Heisenberg one. This is
\begin{equation}\lb{uhr}
\sigma_1^2 \sigma_2^2 \sim \frac{n}{4}.
\end{equation}
It is clear that the absence of correlation between lowering and
raising operators leads to the Heisenberg uncertainty relation. Note
that, one can verify that a similar result holds for the Landau
system on two-sphere in the presence of a strong magnetic
field. Furthermore, in this limit, one can obtain a
two-dimensional non-commutative plane and think about the fuzzy
spaces. It seems that there is a hidden relation between the
absence of the quantum correlations and the fuzzy structures. In the
case under consideration, LLL provides us
with a realization of the so-called fuzzy disc~\cite{liz}.
Indeed, the relation (\ref{uhr}) suggests that the operators $L_+$ and
$L_-$ can be represented as the harmonic oscillator creation and
annihilation operators.
According to~\cite{liz}, a fuzzy disc can be defined by some
adequate projection on LLL. The projection
operator in~\cite{liz} coincides with the density operator given
by~(\ref{dens}). The algebra of functions on ${\bf B}^1$ reduces
to a non-commutative subalgebra, which is isomorphic to the algebra
of $M\times M$ matrices.
The parameter of non-commutativity is proportional to the strength of
the magnetic field.
The fuzzy disc is endowed with the Voros star product
as expected since for a strong magnetic field the coherent
states~(\ref{cs}) go to those for  the harmonic oscillator and  are
eigenstates of destruction operator.

\section{Conclusion}

The quantum mechanics of a charged particles living on the disc ${\bf
  B}^1$
is analyzed from group theory point of view. This is achieved by
  realizing the disc as the non-compact coset space  $SU(1,1)/
  U(1)$. This realization makes the derivation of the Landau levels
  and the corresponding wavefunctions obvious. The wavefunctions are
  identified to the Wigner ${\cal D}$-functions ${\cal D}_{m',m}^k(g)$
with the condition $m={n\over 2}-k$. The index $k$, labeling the unitary irreducible
  representation of the group  $SU(1,1)$, is related to the
  the strength of the magnetic field.
It is remarkable that the LLL  wavefunctions
  coincide with the   $SU(1,1)$ coherent states. The spectrum of the
  Landau problem on the disc is generated from the  $SU(1,1)$ Casimir
  operator.

Restricting to LLL, we have derived the effective WZW action that
describes the quantum Hall droplet of radius proportional to
$\sqrt{M}$, with $M$ is the number of particles in LLL.
To obtain the action of the boundary
excitations, we have defined the star product and  density of
states. Also we have introduced the perturbation potential
responsible of the degeneracy lifting in terms of the $L_3$ left
generator of $SU(1,1)$. We have analyzed the nature of the edge
excitations. Finally, we have shown that the LLL
wavefunctions minimize the RS uncertainty relation and for a strong
magnetic field reduce to the harmonic oscillators coherent states.
As by product one can define the fuzzy space equipped with the
non-commutative Voros star product which, emerges in this case since
the coherent state is eigenvector of the annihilation operator.

Of course still some questions to be answered. It is natural to
ask about the nature of edge excitations in higher dimensional
spaces. On the other hand, is there some way to deal with QHE on
the flag spaces. These different issues are under consideration.

\section*{Acknowledgment}

MD work's was partially done  during a visit to
the Max Planck Institute for Physics of
Complex Systems (Dresden, Germany). He would like to acknowledge the
financial support of the Institute.

\end{document}